\documentclass[10pt]{article}

\usepackage{hyperref}
\usepackage{amsmath}
\usepackage{graphicx}
\usepackage{subfigure}

\title{Growth kinetics and morphological stability of precipitates in 3-D: a phase field study}

\author{Arijit Roy and M. P. Gururajan \\ Department of Metallurgical Engineering and Mateials Science, \\ Indian Institute of Technology Bombay, Powai, Mumbai 400076 INDIA}

\begin{document}

\maketitle

\section*{Abstract}
We have studied the growth kinetics  of isolated precipitates growing from a supersaturated matrix in 3-dimensions (3-D) using phase field models; we assume isotropic interfacial energy consider both constant and variable diffusivity. We report and compare our numerical growth rates with the classic analytical solutions of Zener and Frank (ZF). The numerical results deviate from the analytical ones. These deviations can be understood in terms of the generalised Gibbs-Thomson effect. Specifically, due to the higher capillary contribution in 3-D (curvature is twice for a sphere compared to a circle), the precipitate growth kinetics deviates more from ZF in 3-D as compared to 2-D. In addition, the kinetic parameter associated with the normal velocity of the precipitate-matrix interface also modifies the deviation of the precipitate composition from its equilibrium value and hence its growth kinetics. In phase field models (such as the one used by us) which use a combination of Allen-Cahn and Cahn-Hilliard type equations, we show how to choose the kinetic parameters (namely, mobility and relaxation parameter) so that the kinetic coefficient (in the generalised Gibbs-Thomson equation) is made effectively zero. We also show that the kinetic parameter the precipitate-matrix interface might play a crucial role in making the precipitate undergo morphological instabilities as it grows (leading to ``sea-weed''-like structures).

\section*{Keywords}
Generalised Gibbs-Thomson effect, Precipitate growth kinetics,  Phase field modelling, Zener-Frank growth kinetics, 3-D precipitate growth kinetics, morphological instabilities



\section{Introduction}

The growth kinetics of precipitates is an important problem in solid-solid phase transformations. The classical results on the growth kinetics of precipitates are due to Zener~\cite{Zener} and Frank~\cite{Frank} (hereinafter, referred to as ZF). Frank, for example, described the relationship between the (normalised) growth rate of an isolated precipitate as a function of (normalised) supersaturation of the matrix from which it is growing, in systems with radial symmetry in 2- and 3-D (circular and spherical, respectively). 

The presence of interfaces alters the free energies. In a binary alloy, the interfacial free energy alters the compositions of the phases in equilibrium across a curved interface (as opposed to a planar interface). This change in equilibrium composition can be related to the curvature of the interface and the interfacial energy. The expression that connects the differences in composition between the planar and curved interfaces to that of the interfacial energy and curvature is known as the Gibbs-Thomson equation~\cite{PorterEasterling,JWChristian,Johnson,Perez}.  In recent times, there have also been several experimental and computational attempts to both evaluate the Gibbs-Thomson effect and to use it to estimate the interfacial energies: see for example~\cite{ShahandehNategh,DuEtAl1,DuEtAl2}. In ZF, however, this well-known effect of interfacial energy (or, ``capillary effect") is not accounted for.

Recently, phase field models have been used quite successfully to study the growth kinetics of precipitates in systems with constant and variable diffusivity in 1- and 2-D systems~\cite{Rajdip1,Rajdip2}. These studies have shown that the phase field models match with the results of Frank and Zener in 1-D. On the other hand, in 2-D, the growth kinetics deviate from (specifically, are lower than) those predicated by ZF and are size-dependent; and, these deviations are due to the Gibbs-Thomson effect. Given this, in 3-D, where the curvature can be more prominent, one can expect to Gibbs-Thomson to play a dominant role in determining the growth kinetics. In this paper, by extending the implementation of the phase field models described in~\cite{Rajdip1,Rajdip2} (with constant and variable diffusivity and isotropic interfacial energy) to 3-D, we show that this indeed is the case.
 
In the kinetic setting, say the growth of a second phase precipitate from a supersaturated matrix, the Gibbs-Thomson effect gives the boundary conditions at the interface. In such a setting, in addition to the classical Gibbs-Thomson effect due to capillarity, there is also the kinetic effect due to the atomistic processes at the interface (attachment kinetics). That is, the boundary condition at the interface is decided both by capillarity and the normal velocity of the interface~\cite{Davis,Elder,DantzigRappaz}.  The form of the Gibbs-Thomson equation that includes both the capillary and interface normal velocity terms is known as the generalised Gibbs-Thomson effect. In the context of growth of crystals from vapour, the kinetic term is described by the Hertz-Knudsen equation~\cite{KaempferPlapp,Libbrecht}. In the context of solidification of a solid from its melt, the kinetic term is known as kinetic undercooling. Experimentally, it is well known that such generalised Gibbs-Thomson can affect nucleation~\cite{NishiokaMaksimov}, growth~\cite{Libbrecht,SociEtAl,JohanssonEtAl,DayehPicraux} and morphology~\cite{LibbrechtRPP}. 

We use a phase field model based only on the Cahn-Hilliard equation for the study of growth kinetics in systems with variable diffusivity. We use a phase field model based on both the Allen-Cahn and Cahn-Hilliard equations for the study of growth kinetics in systems with constant diffusivity. Such combined Allen-Cahn and Cahn-Hilliard models (known as Model C in the classification of Hohenberg and Halperin~\cite{HohenbergHalperin}) have been widely used -- to study precipitate growth kinetics~\cite{Rajdip1,Rajdip2}, Widmanstatten formation~\cite{LoginovaAgrenAmberg} and solidification~\cite{Elder}. Elder et al~\cite{Elder} have carried out thin interface limit studies on these models and have shown that depending on the parameters used in such models, the kinetic coefficient in the generalised Gibbs-Thomson can be either positive or negative (while, in the case of Allen-Cahn, it is always positive, and in the case of Cahn-Hilliard it is always negative). Thus, it is clear that while using Model C, by appropriate choice of kinetic parameters the kinetic coefficient can be made zero. In this paper, we show how, by carrying out a series of 1-D simulations, the choice of parameters can be identified for which the kinetic coefficient can be made negligible. We also show how a limited amount of control can be exercised in the Cahn-Hilliard model on kinetic coefficient by appropriate choice of mobility parameter.

The morphological stability of growing interfaces is also affected by interfacial energy and the kinetic parameter. In the solidification literature, for example, it is known that at small undercooling and high anisotropy leads to dendritic break-up while at higher undercoolings, interface can break-up even if interfacial energy is isotropic~\cite{BrenerEtAl}. The earlier phase field models of precipitate growth kinetics have not shown such instabilities. We show that for the appropriate choice of the kinetic parameters, the precipitates do undergo morphological instabilities.

The rest of this paper is organised as follows: in the next section, we briefly describe the formulations and their numerical implementation using semi-implicit Fourier spectral implementation; in section~\ref{Section3}, we describe some of the salient results from our studies in 3-D systems with constant and variable diffusivity and compare the same with ZF; the results are ratioanlised in terms of the generalised Gibbs-Thomson equation; we also present our results on precipitates that undergo morphological instabilities for appropriate choice of the kinetic coefficient; we conclude the paper in section~\ref{Section4} with a summary of important results.  

\section{Formulations and their numerical implementation}

We have used two different models: Model I (a combination of Allen-Cahn and Cahn-Hilliard equations) for constant diffusivity simulations and Model II (a Cahn-Hilliard equation) for variable diffusivity simulations. Our formulations are identical to that described in~\cite{Rajdip1} (for Model I) and~\cite{Rajdip2} (for Model II) for the growth of an isolated precipitate particle $p$ from supersaturated matrix $m$ in a binary alloy at constant temperature. In addition, we have assumed that there is no elastic contribution to the free energy; this is equivalent to assuming that there are neither coherency strains or nor volume differences between the two phases. In~\cite{Rajdip1,Rajdip2}, the numerical implementation was carried out only for the 1- and 2-D systems; however, here we report on the results from a numerical implementation of the formulation for the 3-D system. For the sake of completion, in this section, we briefly describe the salient features of the models. We refer the interested readers to Ref.~\cite{Rajdip1,Rajdip2} for more detailed information.

\subsection{Formulation: Model I}

The microstructure is described using a combination of a conserved ($c$, the composition) and a non-conserved ($\eta$, the structural order parameter) order parameters. The free energy functional is given by 
\begin{equation}
   F = N_V \int_V[f(c,\eta) + \kappa_c (\nabla c)^2 + \kappa_\eta (\nabla \eta)^2]dV \label{E:F_preci},
\end{equation}
where, $N_V$ is the number of atoms per unit volume, $f(c,\eta)$ is the bulk free energy density, and $\kappa_c$ and $\kappa_\eta$ are  the gradient energy coefficients for $c$ and $\eta$, respectively; the bulk free energy density of the system is given by
\begin{equation}
  f(c,\eta) = A c^2 (1-W(\eta))+B (1-c)^2 W(\eta)+P\eta^2(1-\eta)^2 \label{E:f_preci}
\end{equation}
where, the constant $P$ is used to set the free energy barrier height between the matrix ($m$) and precipitate ($p$) phases; $A$ and $B$ are positive constants corresponding to the free energy of the $m$ and $p$ phase respectively; the $W(\eta)$ is the Wang interpolation function~\cite{WangFunction} that interpolates the free energy between the $m$ and $p$ phases:

$W(\eta) =\begin{cases}
0 \quad \textrm{if} \quad \eta<0 \\ 
\eta^3(1-15\eta+6\eta^2) \quad \textrm{if} \quad 0\leq \eta \leq 1 \\  
1 \quad \textrm{if} \quad \eta>1
\end{cases}$
            
The precipitate growth is governed by the evolution equations corresponding to $c$ (Cahn-Hilliard equation) and $\eta$ (Allen-Cahn equation). 
\begin{eqnarray}
\frac{\partial c}{\partial t} &=& M\nabla^2 \mu_c = M\nabla^2 \frac{\delta (F/N_V)}{\delta c}  \label{E:preci_CH} \\
\frac{\partial \eta}{\partial t} &=& -L \mu_\eta = - L \frac{\delta (F/N_V)}{\delta \eta} \label{E:preci_AC}
\end{eqnarray}
where $M$ and $L$ are the (constant) atomic mobility and relaxation parameter, respectively; the chemical potentials $\mu_c$ and $\mu_\eta$ are the variational derivatives of the free energy with respect to $c$ and $\eta$, respectively.

Thus, we solve the following coupled equations to study the microstructural evolution:
\begin{eqnarray}
\frac{\partial c}{\partial t} &=& M\nabla^2 \left[\frac{\partial f(c,\eta)}{\partial c} - \kappa_c \nabla ^2 c \right] \label{E:preci_CH-mod} \\
\frac{\partial \eta}{\partial t} &=& -L \left[ \frac{\partial f(c,\eta)}{\partial \eta} - \kappa_\eta \nabla ^2 \eta \right] \label{E:preci_AC-mod}
\end{eqnarray}

\subsection{Formulation: Model II}

The microstructure is described using the conserved order parameter $c$, the composition. The free energy functional is given by 
\begin{equation}
   F = N_V \int_V[f(c) + \kappa (\nabla c)^2]dV \label{CHFreeEnergy},
\end{equation}
where, $N_V$ is the number of atoms per unit volume, $f(c)$ is the bulk free energy density, and $\kappa$ and is the gradient energy coefficient; the bulk free energy density of the system is given by
\begin{equation}
  f(c) =  A_c c^2 (1-c)^2 \label{CHfZero}
\end{equation}
where, the constant $A_c$ is used to set the free energy barrier height between the matrix ($m$) and precipitate ($p$) phases.
            
The precipitate growth is governed by the evolution equation corresponding to $c$ (Cahn-Hilliard equation):
\begin{equation}
\frac{\partial c}{\partial t} = M_c \nabla^2 \mu = M_c \nabla^2 \frac{\delta (F/N_V)}{\delta c} \label{CahnHilliard}
\end{equation}
where $M_c$ is the (constant) atomic mobility; the chemical potential $\mu$ is the variational derivative of the free energy with respect to $c$.

Thus, we solve the following equation to study the microstructural evolution:
\begin{equation}
\frac{\partial c}{\partial t} = M_c\nabla^2 \left[\frac{\partial f(c)}{\partial c} - \kappa \nabla ^2 c \right]  \label{CHEvolEquation}
\end{equation}

\subsection{Numerical Implementation} 

We have implemented semi-implicit Fourier spectral method to solve Eq.~\ref{E:preci_CH-mod} and ~\ref{E:preci_AC-mod} for Model I and Eq.~\ref{CHEvolEquation} for Model II. In this numerical scheme, the spatial Fourier transforms are carried out for the order parameters and the equations are solved in the reciprocal space; the linear terms are evaluated implicitly while the non-linear terms are evaluated explicitly; the temporal derivatives are discretized using Euler forward difference scheme.

In the following, a variable with a tilde denotes the Fourier transformation for the given quantity, and, the boldfaced letters denote vector quantities.

In our numerical implementation, the required discrete Fourier transforms were carried out using FFTW~\cite{FFTW}. 

The typical system sizes used in the studies are as follows: 20000 (in 1-D), 2048 $\times$ 2048 (in 2-D), and 512 $\times$ 512 $\times$ 512 (in 3-D). 

\subsubsection{Model I}

The evolution equations that are solved for Model I are as follows: 
\begin{equation}
 \tilde{c}(\textbf{k},t+\Delta t) = \frac{\tilde{c}(\textbf{k},t) - k^2\Delta t M \tilde{g}_c(\textbf{k},t)}{1+2\Delta t M \kappa_c k^4} \label{E:preci_numC}
\end{equation}

\begin{equation}
 \tilde{\eta}(\textbf{k},t+\Delta t) = \frac{\tilde{\eta}(\textbf{k},t) - \Delta t L \tilde{g}_\eta(\textbf{k},t)}{1+2\Delta t L \kappa_\eta k^2} \label{E:preci_numE}
\end{equation}

where, $\textbf{k}$ is the wave vector ($= {2\pi}/L$) with $k=|\textbf{k}|$ and
\begin{equation}
 \tilde{g}_c = \widetilde{\frac{\partial f(c,\eta)}{\partial c}}; \;\;\;\;  \tilde{g}_\eta = \widetilde{\frac{\partial f(c,\eta)}{\partial \eta}}
\end{equation}

We have used non-dimensional values of unity for $P$, $A$, $B$, $\kappa_c$, and $\kappa_{\eta}$; for details of the non-dimensionalisation, we refer the reader to~\cite{Rajdip1}. Most of our growth simulations are carried out using non-dimensional values of $L=1$ and $M = 2.166$ -- the reason for this choice is explained in the next section. The numerical simulations are carried out using the following spatial and temporal discretizations: $\Delta x = \Delta y = \Delta z = 0.4 $; and $\Delta t = 0.2$. 

\subsubsection{Model II}

The evolution equation that is solved for Model I is as follows: 
\begin{equation}
 \tilde{c}(\textbf{k},t+\Delta t) = \frac{\tilde{c}(\textbf{k},t) - k^2\Delta t M \tilde{g}_c(\textbf{k},t)}{1+2\Delta t M_c \kappa k^4} \label{CHNumerical}
\end{equation}
where, $\textbf{k}$ is the wave vector ($= {2\pi}/L$) with $k=|\textbf{k}|$ and
\begin{equation}
 \tilde{g}_c = \widetilde{\frac{\partial f(c,\eta)}{\partial c}}
\end{equation}

We have used non-dimensional values of unity for $M_c$, $A_c$, and $\kappa$; for details of the non-dimensionalisation, we refer the reader to~\cite{Rajdip2}. The numerical simulations are carried out using the following spatial and temporal discretizations: $\Delta x = \Delta y = \Delta z = 0.4 $ (in 2- and 3-D) and $\Delta x = \Delta y = \Delta z = 1.0 $ in 1-D; and $\Delta t= 0.5$ (in 1- and 2-D) and $\Delta t = 0.2$ in 3-D. 

\subsection{Interfacial energy and critical nuclei radius}

Using the 1-D simulations, in which we start with a box initial profile (that is, a precipitate embedded inside a matrix with a very sharp interface) and equilibrate the same (so that the interfaces reach the appropriate profile), we can calculate the interfacial energy $\sigma$ using the following integral for Model I: 
\begin{equation}
 \sigma = \frac{1}{2} \int_0^l (f(c,\eta) + \kappa_c (\nabla c)^2 + \kappa_\eta) (\nabla \eta)^2)dx \label{E:precir_sigma}
\end{equation}
and, the following equation for Model II:
\begin{equation}
 \sigma = \frac{1}{2} \int_0^l (f(c) + \kappa (\nabla c)^2 dx \label{InterfacilEnergyCH}
\end{equation}
where $l$ is the length of the simulation cell. For our chosen parameters, the non-dimensional interfacial energy in our system is 0.97 in Model I and 0.33 in Model II. 

Using the interfacial energy, and using the driving force for nucleation (free energy per unit volume, $\Delta G_v$) for the given supersaturation of $c_\infty$ -- calculated using the equation~\cite{PorterEasterling}
\begin{equation}
 \Delta G_v (c_\infty) = - A (c_\infty)^2 - 2 A c_\infty (1-c_\infty) \label{E:precir_DeltaG_v}
\end{equation}
we have calculated the critical nuclei radii for different far-field compositions in 2- and 3-D systems. In our simulations, the initial particle sizes are chosen to be slightly bigger than this critical radii. 

\section{Results and discussion} \label{Section3}

The Gibbs-Thomson effect induces shifts in precipitate composition -- due both to curvature (through interfacial energy $\sigma$) and velocity of the interface (through the kinetic parameter $\beta$). If $\Delta c$ is the change in composition,
\begin{equation}
\Delta c = \frac{\sigma K}{2} + \beta v
\end{equation}
where $K$ is the curvature, and $v$ is the normal velocity of the interface.

\subsection{On the choice of kinetic model parameters}

For Model C (Model I in our terminology), Elder et al have shown (using thin interface limit asymptotics) that the kinetic parameter $\beta$ depends on the mobility $M$ and attachment kinetics $L$ as follows:
\begin{equation} \label{betaMLequation}
\beta = C_0  \left[ \frac{C_1}{L} - \frac{C_2}{M} \right]
\end{equation}
where $C_0$, $C_1$ and $C_2$ are constants that depend on the phase diagram (equilibrium compositions) and the form and parameters of free energy functional.

\begin{figure}[htbp]
\centering
\includegraphics[height=2.5in,width=2.5in]{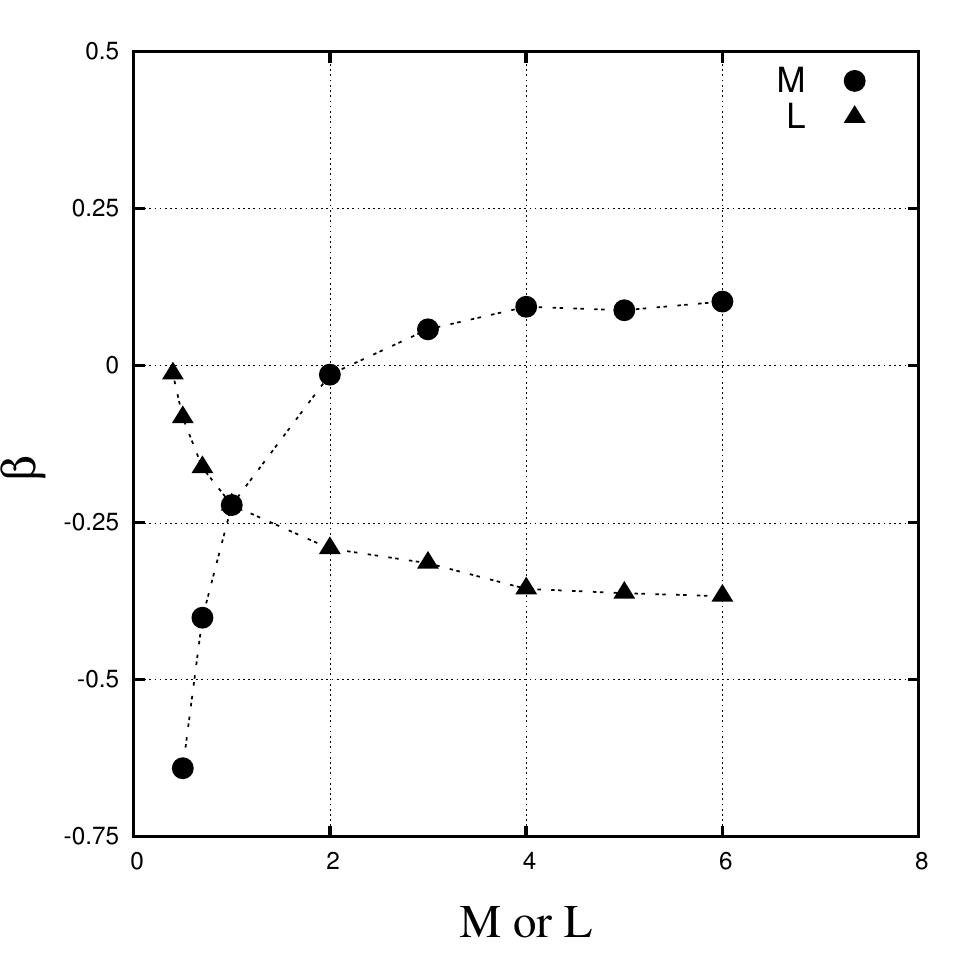}
\caption{Model I: The effect of the kinetic parameters (M and L) on the kinetic coefficient $\beta$. When $M$ varies $L$ is kept at unity and vice versa. In this figure, when $M$ is varied $L$ is kept at unity (and vice versa). Note that the lines joining the data points are only a guide to the eye.}\label{MLvsBeta}
\end{figure}

We have used 1-D simulations to calculate the $\beta$ for the given combination of $M$ and $L$ parameters. We grow the precipitates (in 1D) for a given supersaturation (of, say, $c_{\infty} = 0.4$; note that the $\beta$ value is independent of the supersaturation and our numerical simulations have confirmed this). Since in 1-D, there is no curvature effect, any deviation of the precipitate from the equilibrium composition can be attributed to the interface normal velocity. Thus, from the slope of velocity versus composition deviation, $\beta$ can be estimated. The results of these simulations are summarised in Fig.~\ref{MLvsBeta}. The two curves correspond to varying $M$ (while holding $L$ a constant at unity) and $L$ (while holding $M$ a constant at unity). As is clear from Eq.~\ref{betaMLequation}, when $M$ and $L$ are increased (from one to higher values), $\beta$ saturates to either a positive value (for very high $M$) that is closer to $C_0 C_1/L$, or a negative value (for very high $L$) that is closer to $-C_0 C_2/M$. Our numerical results are in agreement with this trend. Fig.~\ref{MLvsBeta} also shows that for $L=1$, a value of $M = 2.166$ makes the $\beta \approx 0$ ($\beta = 0.0016$, to be specific); in all our results reported in this paper, for the constant diffusivity case, unless specified otherwise, we have used this value. Hence, in these simulations, the contribution of $\beta$ to the deviations in composition can be neglected. In contrast, the values used in~\cite{Rajdip1} leads to negative $\beta$ values of $-0.222$ (for Model I). As we show below, at relatively higher supersaturations, $\beta$ affects the growth kinetics (albeit very mildly). 

\begin{figure}[htbp]
\centering 
\includegraphics[height=2.5in,width=2.5in]{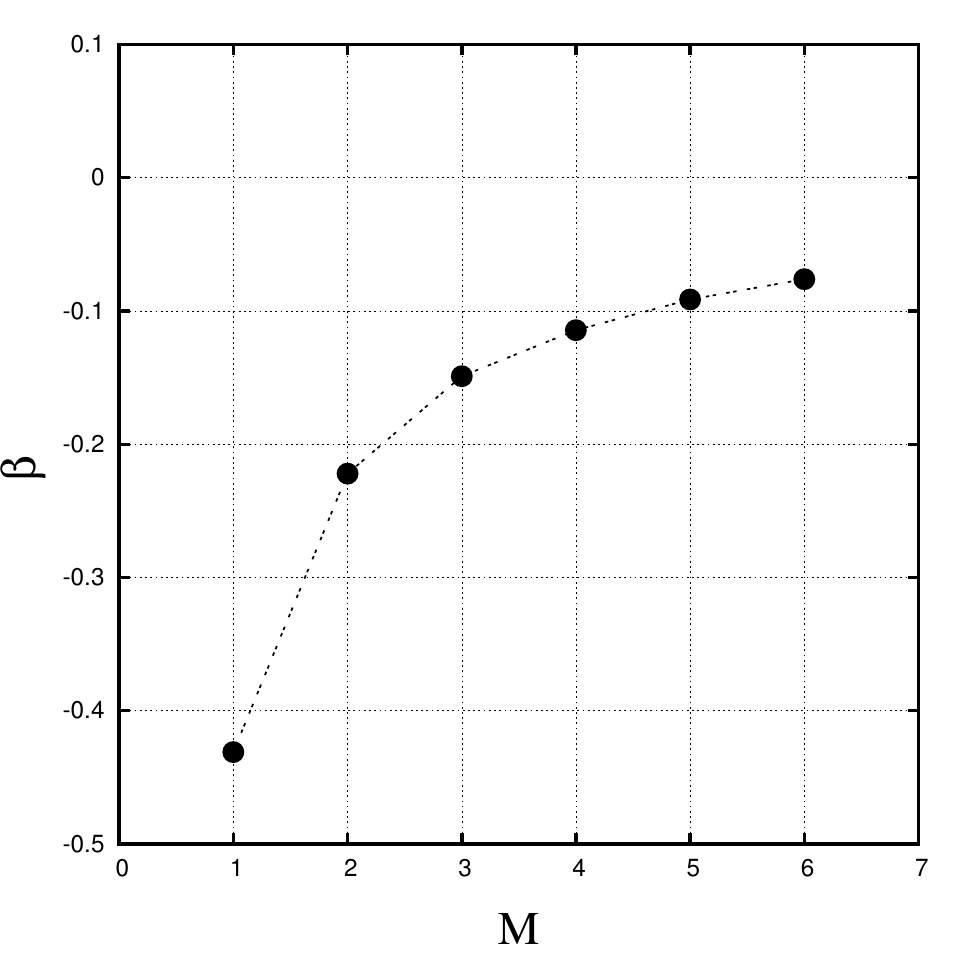}
\caption{Model II: The effect of mobility $M$ on the kinetic coefficient $\beta$. Note that the line joining the data points is only a guide to the eye.}\label{MvsBeta}
\end{figure}

On the other hand, for Model II, we have used a value of $M=1$ which leads to  $\beta = -0.431$ (and is the same as~\cite{Rajdip2}). However, even in this case, it is possible to reduce the magnitude of $\beta$ by increasing $M$. This is shown in Fig.~\ref{MvsBeta}. However, as shown by Elder et al~\cite{Elder}, the sign of $\beta$ remains negative. 

\subsection{Growth kinetics}

The ZF theory has shown that the square of the radius of a growing precipitate ($R^2$) is proportional to $Dt$, where $D$ is the diffusivity and $t$ is the time; the proportionality constant is known as the growth coefficient ($\alpha$):
\begin{equation}
 (R^2 - R^2_0) = \alpha^2 D(t-t_0) \label{E:preicir_Zener}
\end{equation}
where, $R_0$ is the initial radius of the precipitates at time $t_0$. Hence, to calculate the growth rates from our numerical results, we calculate the radius $R$ of the particle at every time $t$ and plot the $R^2$ versus $t$ curves; then, using the expression
\begin{equation}
 \alpha = \sqrt{\frac{R^2(t+\Delta t) - R^2(t-\Delta t)}{2D\Delta t}} \label{E:precir_alphag}
\end{equation}
where $\Delta t$ is the time interval, we calculate $\alpha$.

The ZF theory is valid only for an isolated precipitate growing from a supersaturated matrix. However, in our numerical simulations, we have used periodic boundary conditions. Hence, in all the results we report below, we have discarded the data points when the far-field composition at the edges of the simulation cell changes by about 1\% or more. 

In Model II, the composition dependence of the diffusivity of the matrix phase is given by the expression $D(c) = 2M (6 c^2-6c+1)$~\cite{Rajdip2}. Thus, the diffusivity becomes negative when the matrix composition reaches the spinodal point $c_s \approx 0.2113$. Hence, we have restricted our simulations to far-field compositions of $c_{\infty} = 0.1$ and $c_{\infty} = 0.2$ in this case. On the other hand, for Model I, we have used (typically) far-field compositions of $c_{\infty} = 0.1,0.2,0.3, \; {\mathrm{and}} \;0.4$. 

\begin{figure}[htbp]
\centering
\includegraphics[height=2.0in,width=2.0in]{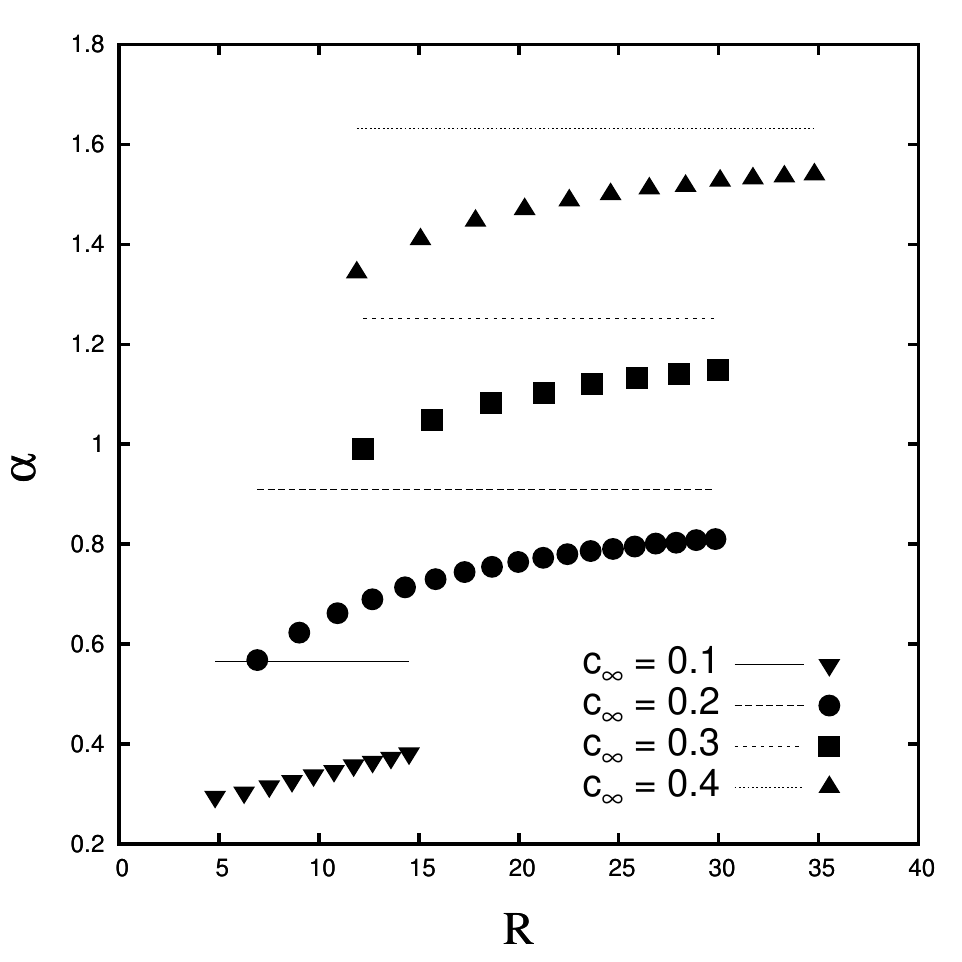}
\includegraphics[height=2.0in,width=2.0in]{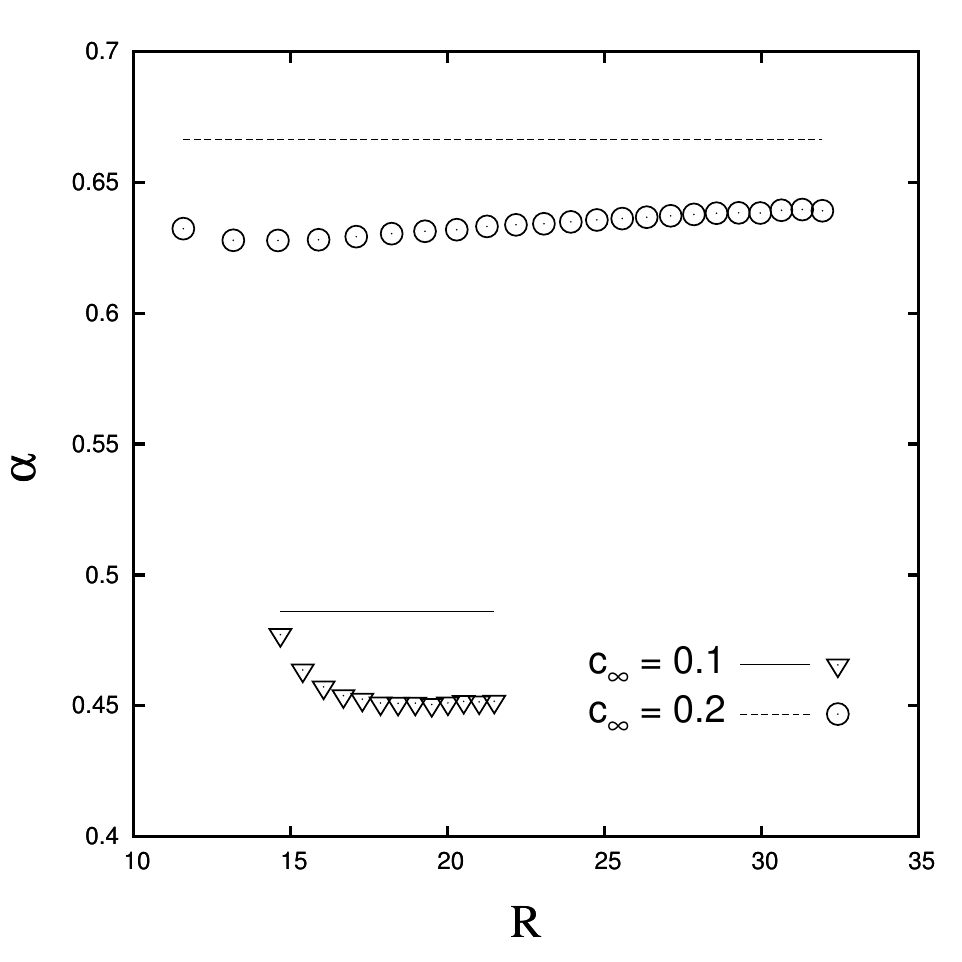}
\caption{The radii of the precipitates ($R$) versus the growth rate ($\alpha$) for constant (left) and variable (right) diffusivities. The data from the numerical simulations are shown by points while lines correspond to the analytical values calculated from ZF.}\label{F:3DRvsAlpha}
\end{figure}

In Fig.~\ref{F:3DRvsAlpha} we show the $\alpha$ calculated from our numerical simulations as a function of the radius R of the precipitate. Since in these cases the shapes of the precipitates remain spherical, we calculated the radius of the precipitate along the three principal axes $x$, $y$, and $z$ and report the average of these values as $R$; the interface is identified typically with a composition of $c=0.5$ (unless specified otherwise). In the figure, we have also marked the analytical $\alpha$ calculated from ZF theory. As is clear from the figure, the numerically calculated $\alpha$ values keep changing with $R$ and reach a relatively constant value at high radii. 

\begin{figure}[htbp]
\centering
\includegraphics[height=2.0in,width=2.0in]{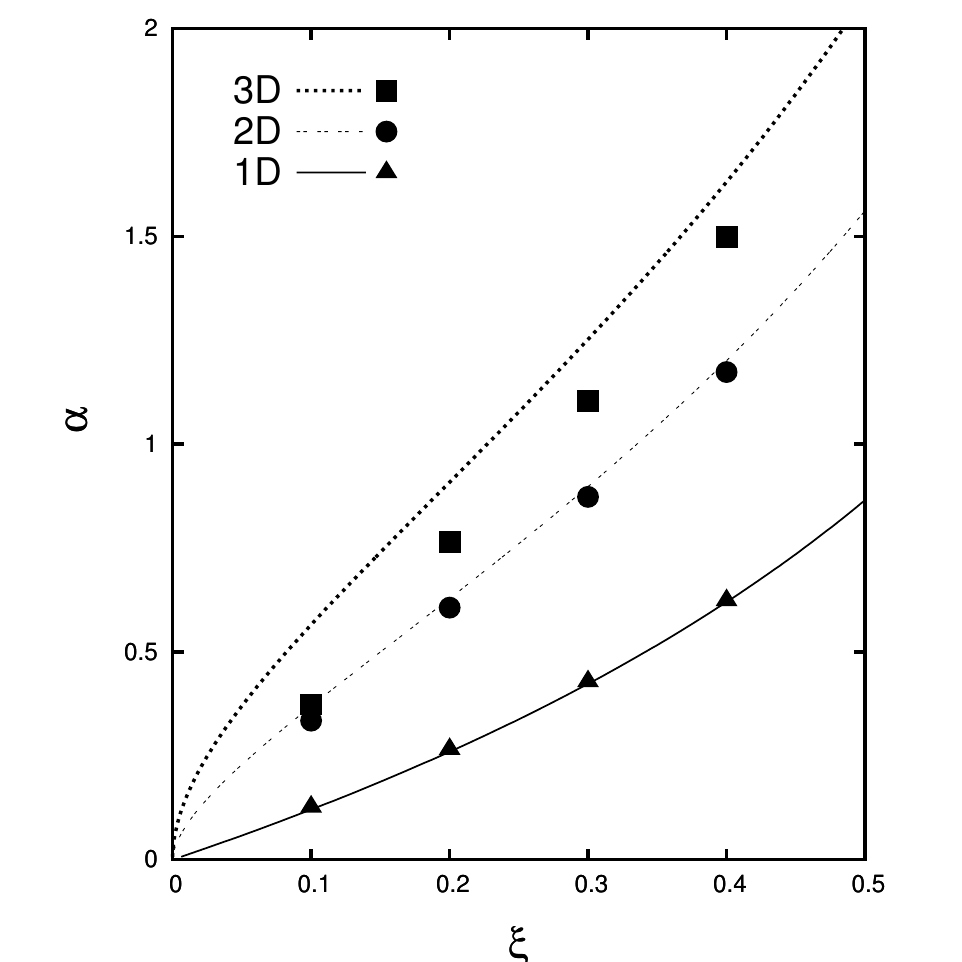}
\includegraphics[height=2.0in,width=2.0in]{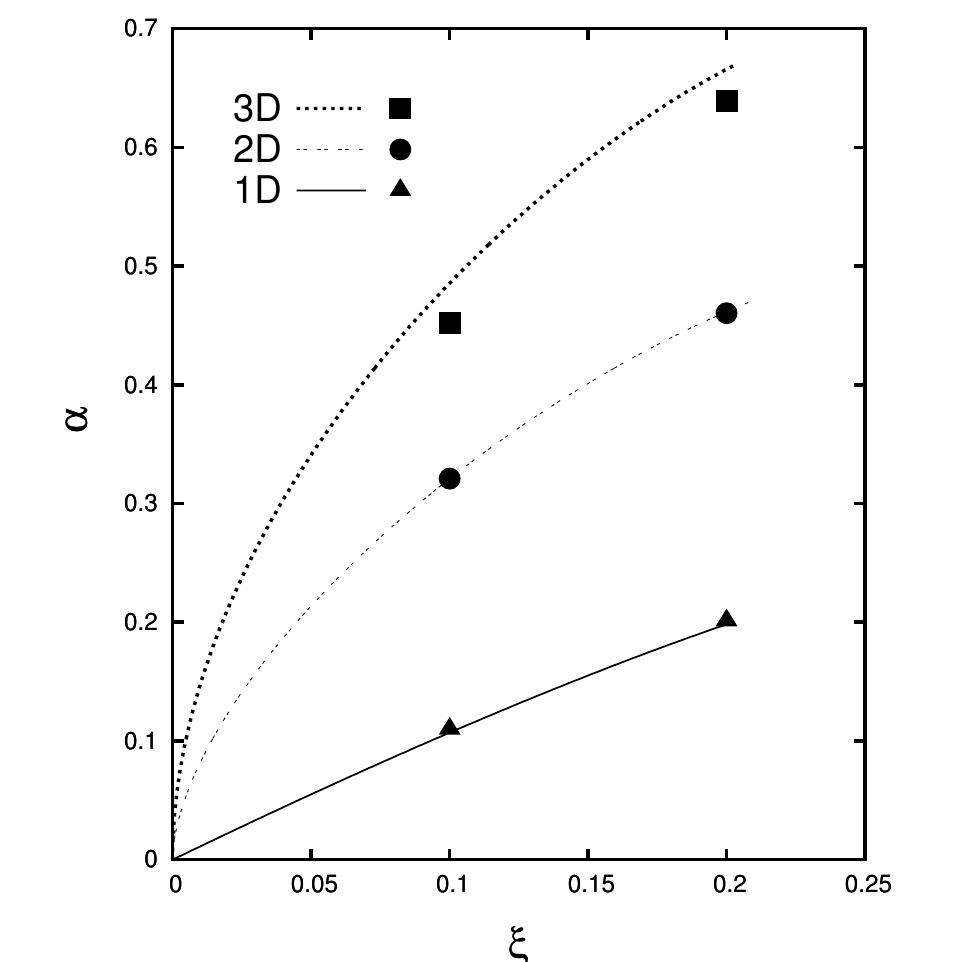}
\caption{The normalised supersaturation $\xi$ versus growth coefficient $\alpha$ for systems with cosntant (left) and variable (right) diffusivity. The data from the numerical simulations are shown by points while lines correspond to the solution of ZF (evaluated analytically for Model I and numerically for Model II). }\label{F:preci_Frankana_PF}
\end{figure}

In Fig.~\ref{F:preci_Frankana_PF}, we show the (normalised) 
supersaturation $\xi$ (defined as $\xi = ({c_\infty - c^m})/({c^p - c^m})$) and the growth coefficient $\alpha$ for 
1-, 2- and 3-D systems, for Model I and Model II, on the left and right, respectively; these $\alpha$ values correspond to the large $R$ values shown in Fig.~\ref{F:3DRvsAlpha}. 
The solid lines correspond to the the analytical expressions derived by  Frank~\cite{Frank}; for Model II, these expressions have been calculated numerically~\cite{RajdipThesis}; for Model I, they are calculated using Eq.~\ref{E:ZF_1D} for 1-D, Eq.~\ref{E:ZF_2D} for 2-D, and Eq.~\ref{E:ZF_3D} for 3-D systems:
\begin{equation}
\xi(\alpha)=\frac{\alpha}{2} \exp\left(\frac{\alpha^2}{4} \right)\pi^{\frac{1}{2}} \left[1-\mathrm{erf} \left(\frac{\alpha}{2} \right) \right]\label{E:ZF_1D}
\end{equation}
\begin{equation}
\xi(\alpha)=\frac{\alpha^2}{2} \exp\left(\frac{\alpha^2}{4}\right) \left \{-\frac{1}{2} \mathrm{Ei}\left(-\frac{\alpha^2}{4} \right) \right\}\label{E:ZF_2D}
\end{equation}
\begin{equation}
\xi(\alpha)=\frac{\alpha^3}{2} \exp\left(\frac{\alpha^2}{4}\right) \left\{\alpha^{-1}\exp\left(-\frac{\alpha^2}{4}\right)-\frac{\pi^{\frac{1}{2}}}{2} \left[1- \mathrm{erf} \left(\frac{\alpha}{2} \right) \right] \right\}\label{E:ZF_3D}
\end{equation}
where $\mathrm{erf}$ is the error function and $\mathrm{Ei}$ is the exponential integral. We have used the GNU Scientific Library function calls~\cite{GSL} for evaluating these integrals.

The 1- and 2-D results in Fig.~\ref{F:preci_Frankana_PF} are the same (except for the use of different kinetic parameters) as given in Ref.~\cite{Rajdip1}; the 3-D results are new. As predicted by ZF, the growth rates in 3-D are higher than that of 2-D. However, as compared to 2-D, the deviation from predicted growth rates is much larger for 3-D. These deviations can be understood in terms of the generalised Gibbs-Thomson effect as follows: purely in terms of the curvature term, in 3-D, the curvature contribution is (for a sphere) is twice as large as that in 2-D (for a circle). Hence, in general, the numerical results in 3-D deviate more than those in 2-D from ZF. Having said that, we see that in the case of Model II, the results agree relatively better with ZF than Model I. This is because, the interfacial energy itself is one-third that of constant diffusivity case. Hence, the effect of capillary driven Gibbs-Thomson is relatively weak. On top of it, in this case the kinetic coefficient is negative. Hence, it tends to reduce the Gibbs-Thomson driven composition deviations, which, in turn, leads to better agreement with ZF. 

\begin{figure}[htbp]
\centering
\includegraphics[height=2.0in,width=2.0in]{Figures/M2-166_growh_rate.pdf}
\includegraphics[height=2.0in,width=2.0in]{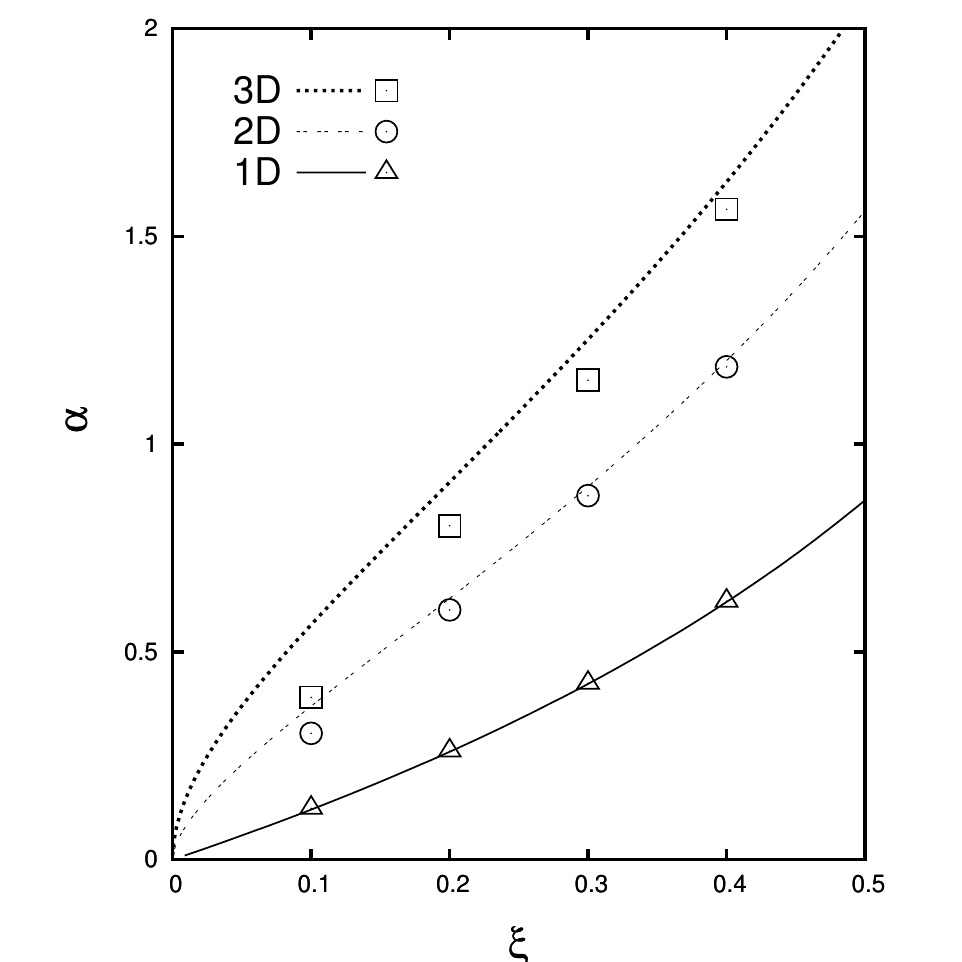}
\caption{The normalised supersaturation $\xi$ versus growth coefficient $\alpha$ for systems with constant diffusivity when $M=2.166$ (left) and $M=1$ (right); these $M$ values correspond to $\beta$ values of $-0.222$ and $0.0016$ respectively. The data from the numerical simulations are shown by points while lines correspond to the analytical solution of ZF. }\label{MComparison}
\end{figure}

In addition, for Model I, in our case, the agreement between numerical results and ZF does not improve with increasing $c_{\infty}$ as was seen by Rajdip et al~\cite{Rajdip1}  -- see Fig.~\ref{MComparison}. When a value of $M=1$ is used (leading to a $\beta$ value of $-0.222$), at higher supersaturations, where the velocities are relatively higher, the agreement with ZF is relatively better than when $\beta \approx 0$. Thus, the kinetic coefficient does play a role in the precipitate growth kinetics, especially at relatively high growth rates.

\begin{figure}[htbp]
\centering
\includegraphics[height=2.0in,width=2.0in]{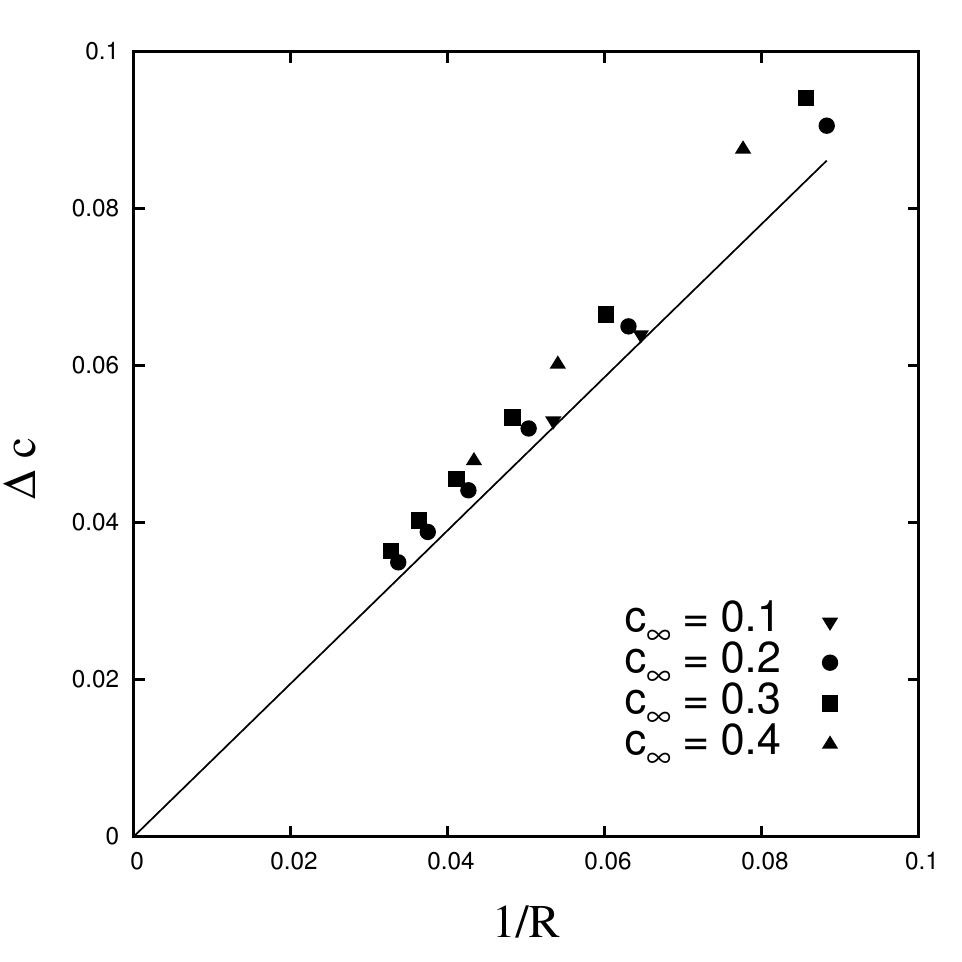}
\includegraphics[height=2.0in,width=2.0in]{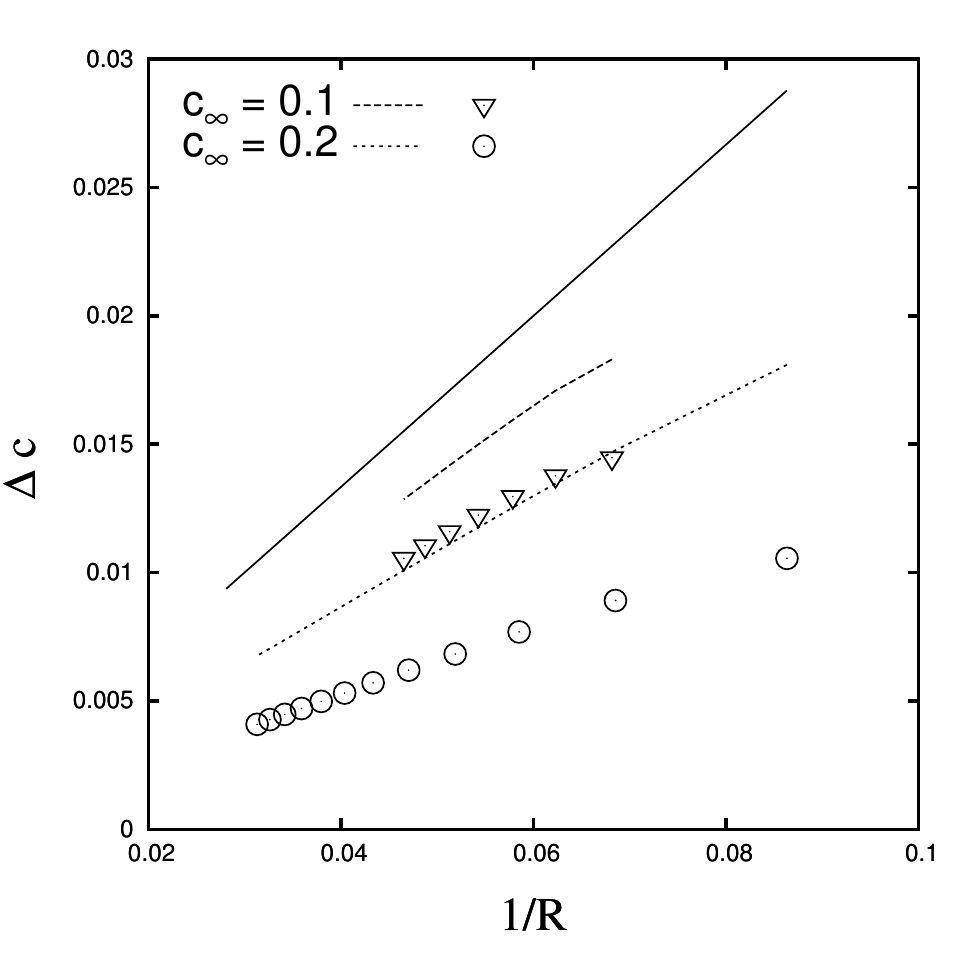}
\caption{The deviation in composition $\Delta c$ versus  $\frac{1}{R}$ of the growing precipitates for constant (left) and variable (right) diffusivity. The data from the numerical simulations are shown by points while lines correspond to the analytical solution. Note that in the case of constant diffusivity, since the classical and generalised Gibbs-Thomson give more or less the same $\Delta c$ (because $\beta$ is negligible), we have only shown the classical Gibbs-Thomson line. In the case of variable diffusivity, however, we show the classical Gibbs-Thomson (continuous line) as well as generalised Gibbs-Thomson (two broken lines corresponding to the two far-field compositions). }\label{F:preci_GT}
\end{figure}

Within the 3-D results, the deviation for smaller $c_{\infty}$ is much larger than for larger $c_{\infty}$. This is because for smaller $c_{\infty}$, the growth rates are smaller and hence, the precipitates achieves $\Delta c$ that is quite close to the analytically predicted $\Delta c$; however, for larger $c_{\infty}$, the numerically obtained $\Delta c$ deviates from that predicated analytically. In Fig.~\ref{F:preci_GT}, we compare the analytically calculated deviation in composition $\Delta c$~\cite{Johnson_Gibbs-Thomson-87}, namely, $\Delta c = \sigma/R$ (where $\sigma$ is the interfacial free energy and $R$ is the precipitate radius) with that numerically obtained from the composition profiles for both constant and variable diffusivity cases. In the case of constant diffusivity (Model I), since we have made $\beta \approx 0$, the classical Gibbs-Thomson and the generalised Gibbs-Thomson give almost the same $\Delta c$. On the other hand, in the case of variable diffusivity (Model II), we show both the classical (continuous line) and generalised Gibbs-Thomson (two broken lines which are calculated by adding the normal velocity of the interface at the given R multiplied by the kinetic coefficient to the classical Gibbs-Thomson). Since the kinetic coefficient is negative, the analytically calculated values are smaller; further since the systems with higher supersaturation grow at higher velocities, the curve for $c_{\infty} = 0.2$ is lower than than for $c_{\infty} = 0.1$. Note that, interestingly, the sign of $\beta$ (even when it is very small) determines whether the numerical $\Delta c$ values lie above (positive $\beta$) of below (negative $\beta$) the analytical curves.
 
\subsection{Morphological instabilities of growing precipitates}

\begin{figure}[htbp]
\centering
\includegraphics[height=1.5in,width=1.5in,trim=2cm 2cm 2cm 2cm]{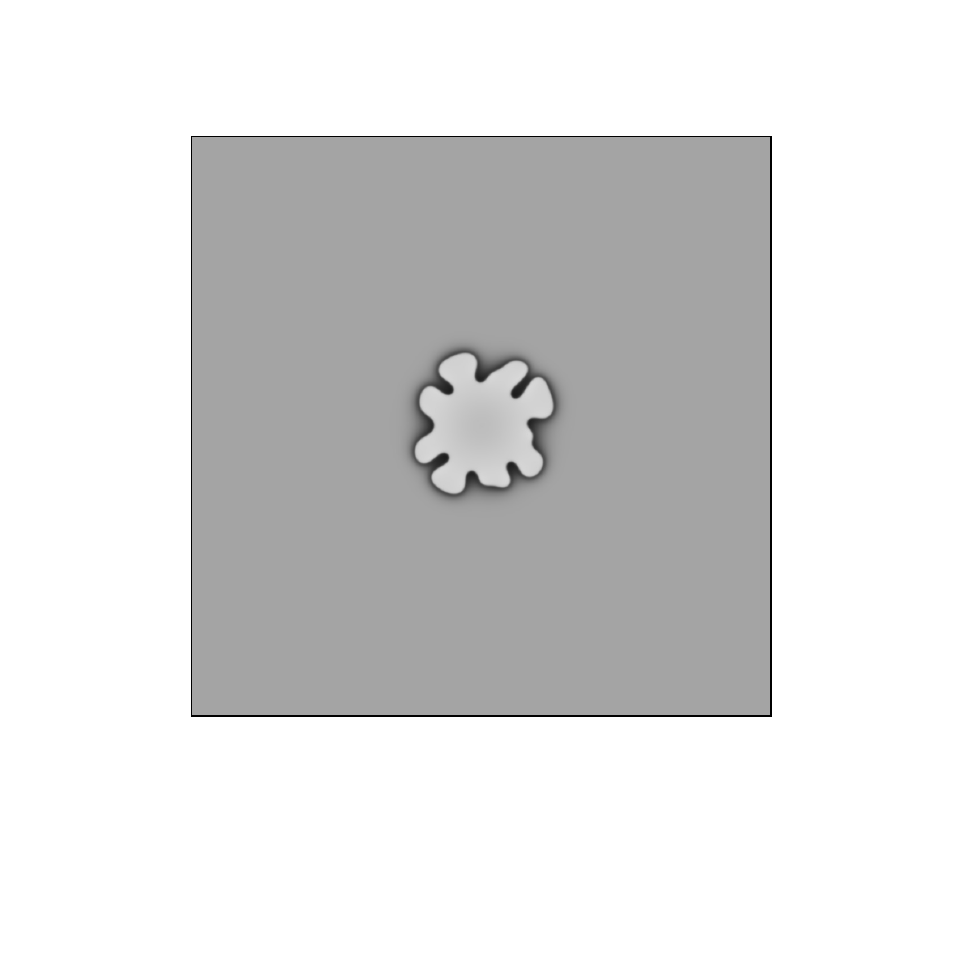}
\includegraphics[height=1.5in,width=1.5in,trim=2cm 2cm 2cm 2cm]{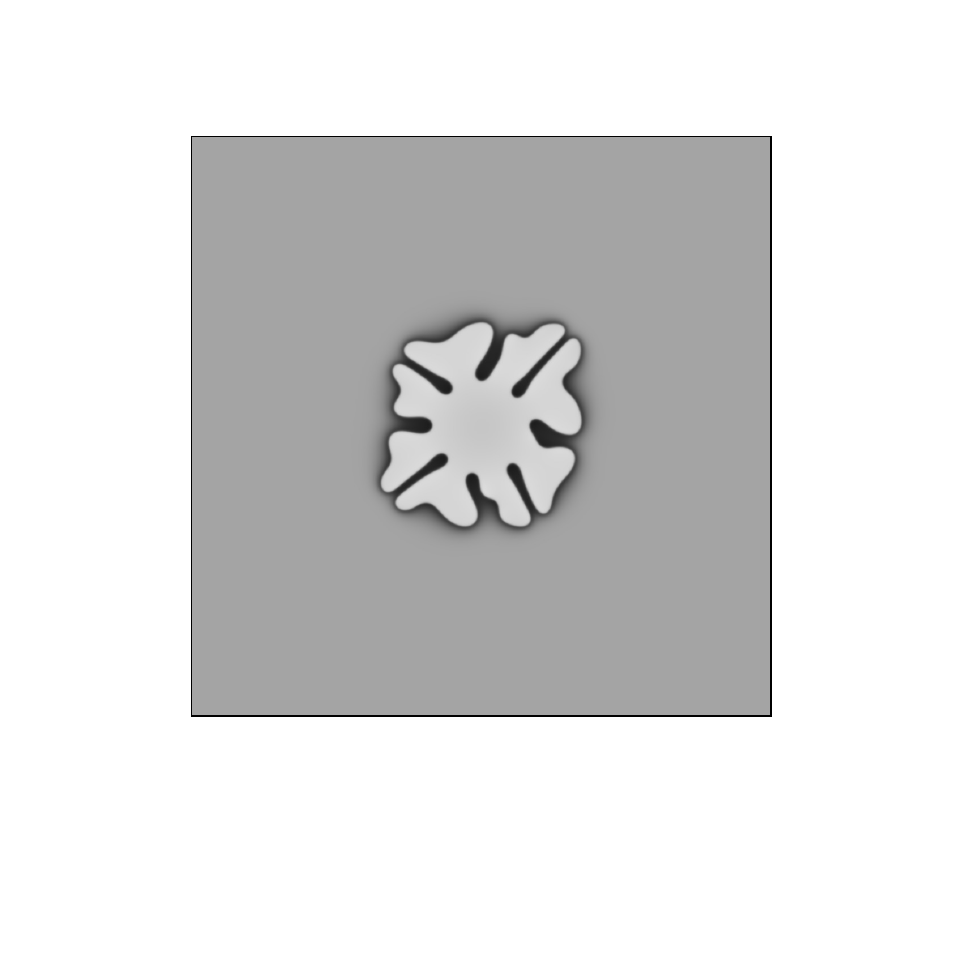}  \\
\includegraphics[height=1.5in,width=1.5in,trim=2cm 2cm 2cm 2cm]{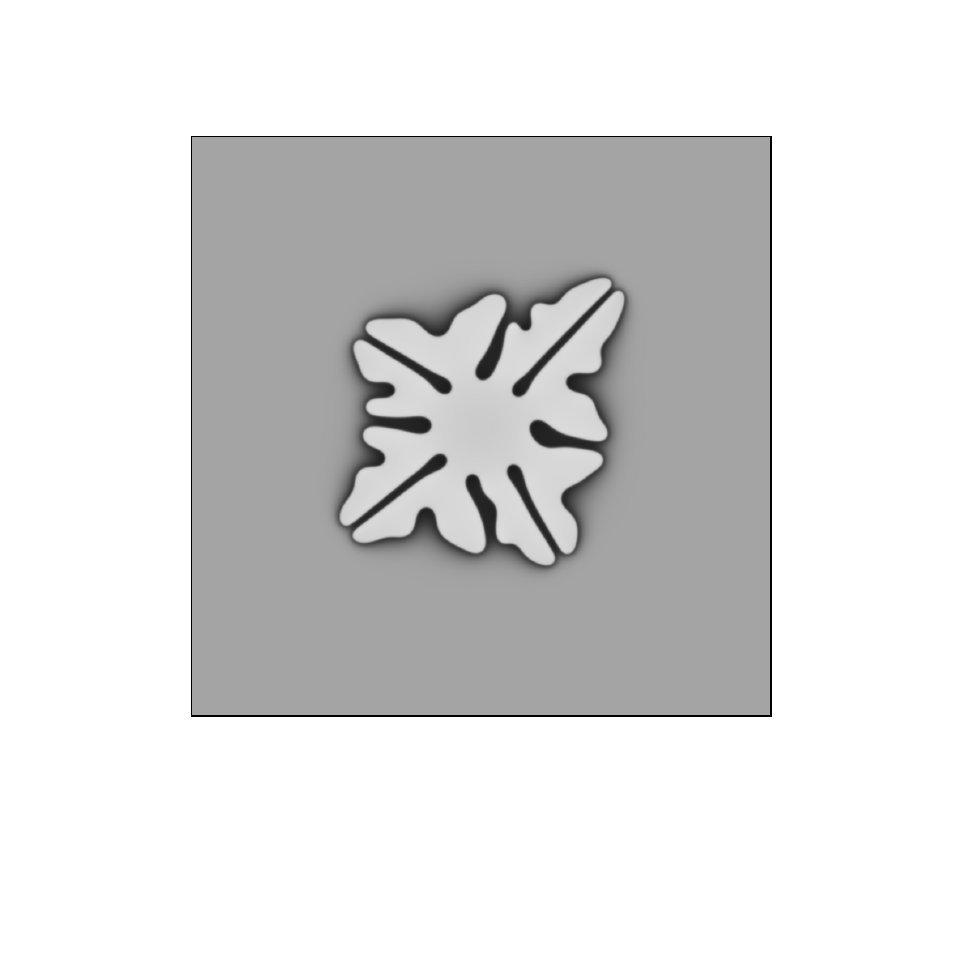}
\includegraphics[height=1.5in,width=1.5in,trim=2cm 2cm 2cm 2cm]{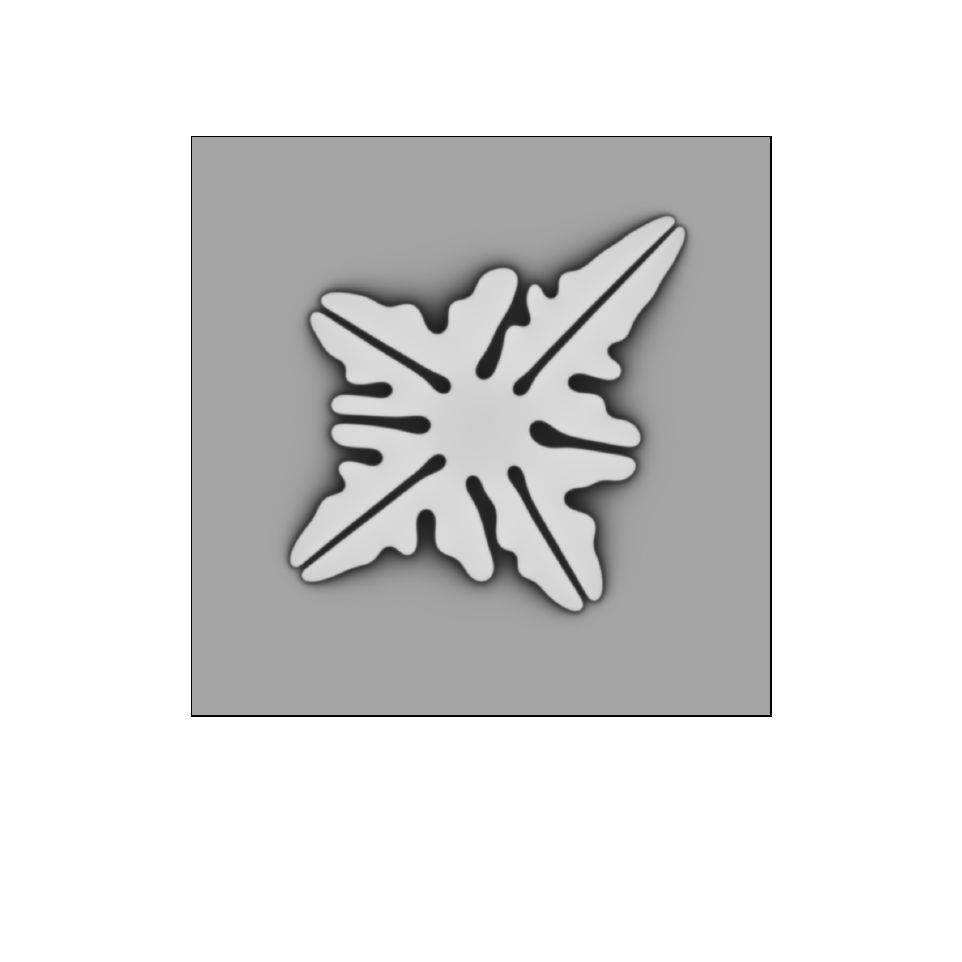}
\caption{The morphological instability of a precipitate growing from a very high supersaturation ($c_{\infty} = 0.7$) and hence at a very high velocity. The microstructures correspond to non-dimensional times of 2000 (top left), 4000 (top right), 6000 (bottom left) and 8000 (bottom right). The initial precipitate was circular. The systems size is 2048 $\times$ 2048. }\label{MorphInstability}
\end{figure}

\begin{figure}[htbp]
\centering
\includegraphics[height=1.0in,width=2.0in,trim=2cm 0cm 2cm 0cm]{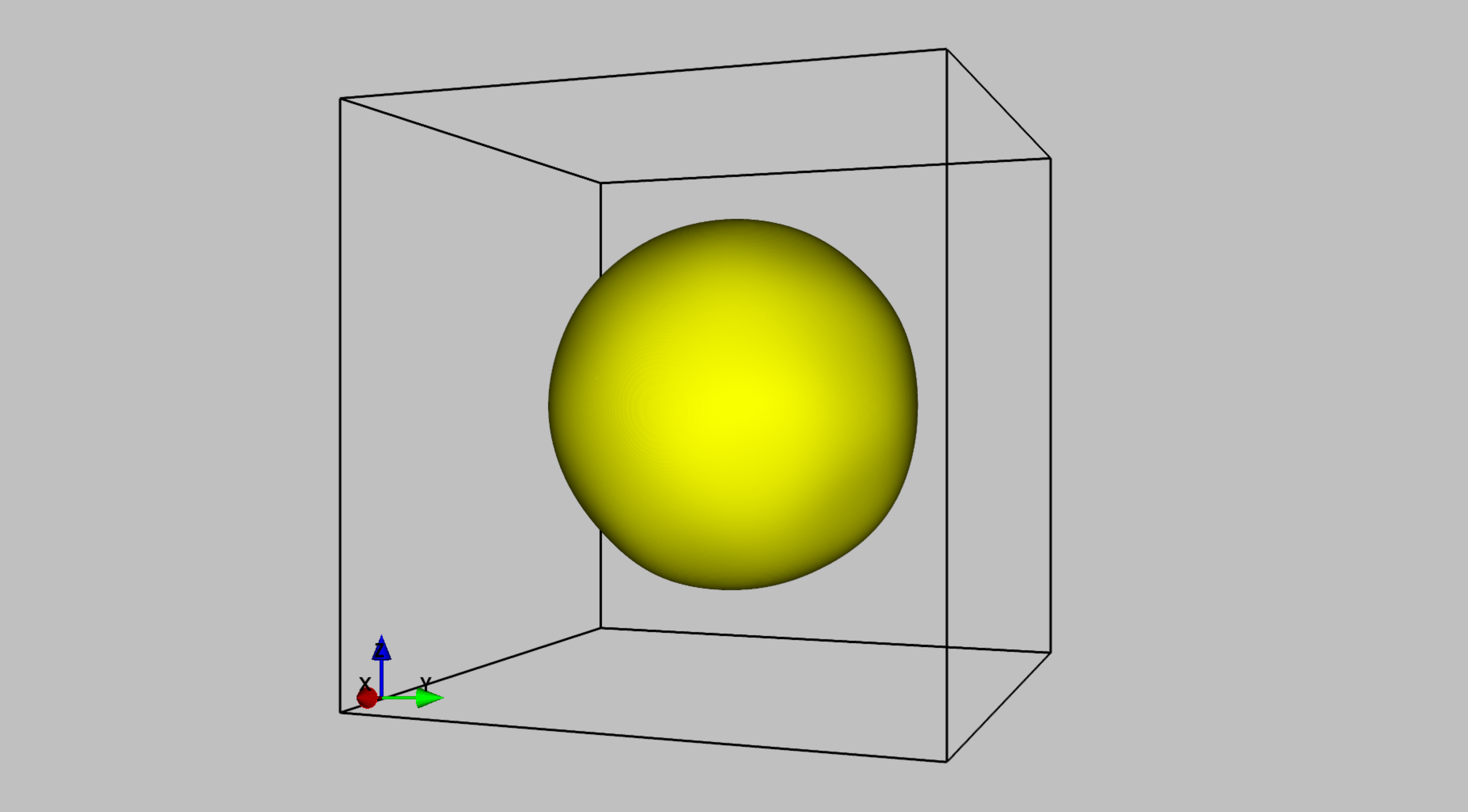}
\includegraphics[height=1.0in,width=2.0in,trim=2cm 0cm 2cm 0cm]{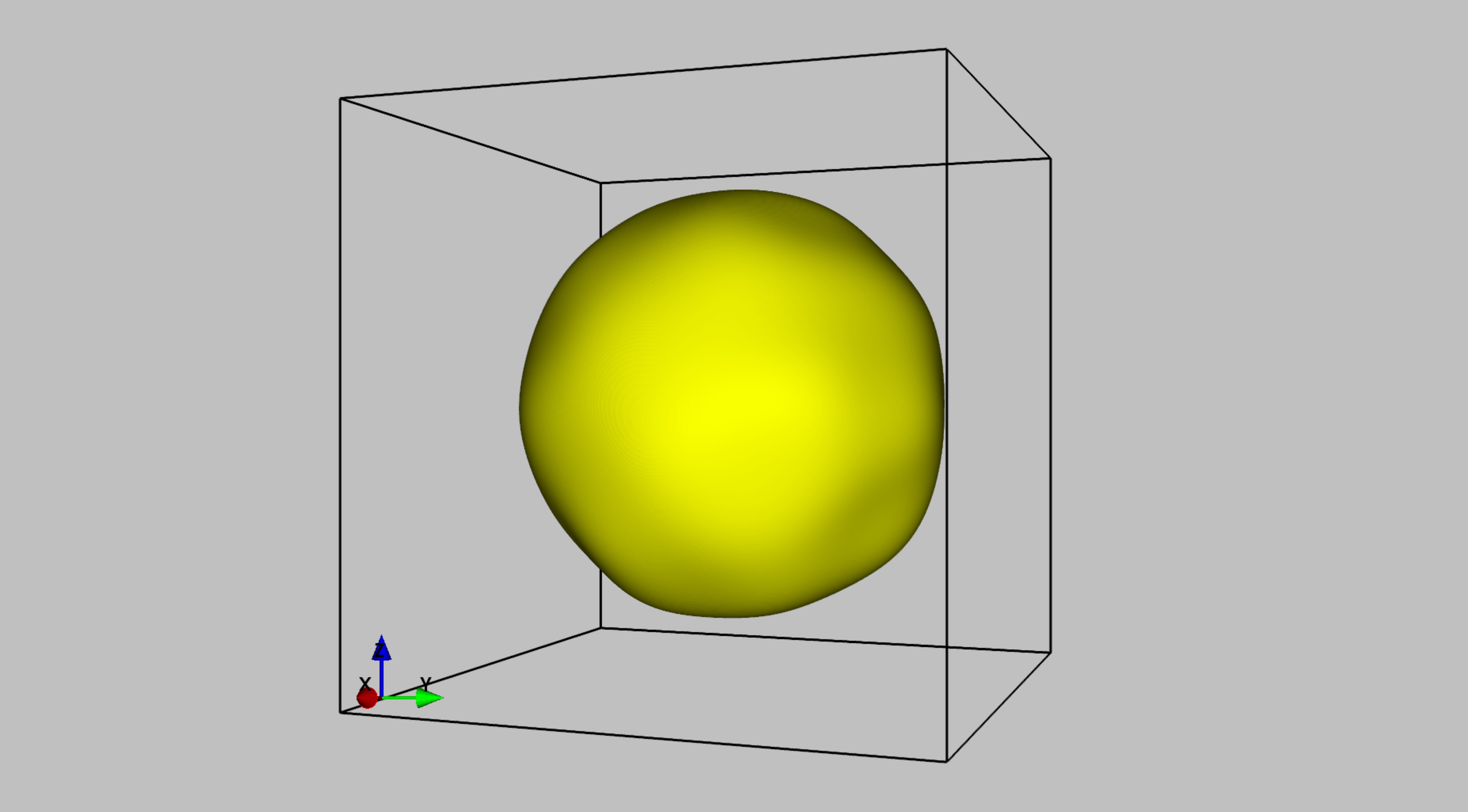}  \\
\includegraphics[height=1.0in,width=2.0in,trim=2cm 0cm 2cm 0cm]{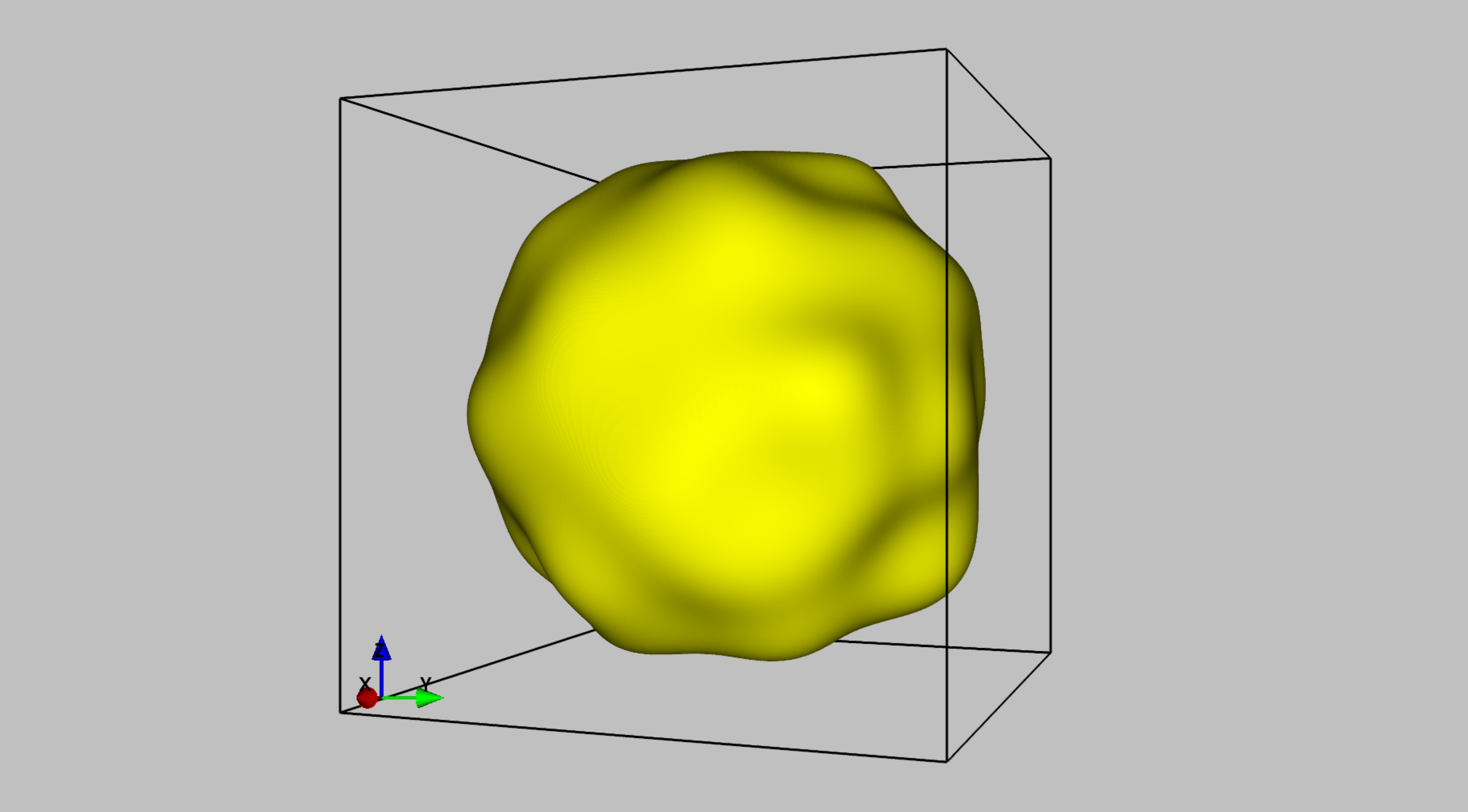}
\includegraphics[height=1.0in,width=2.0in,trim=2cm 0cm 2cm 0cm]{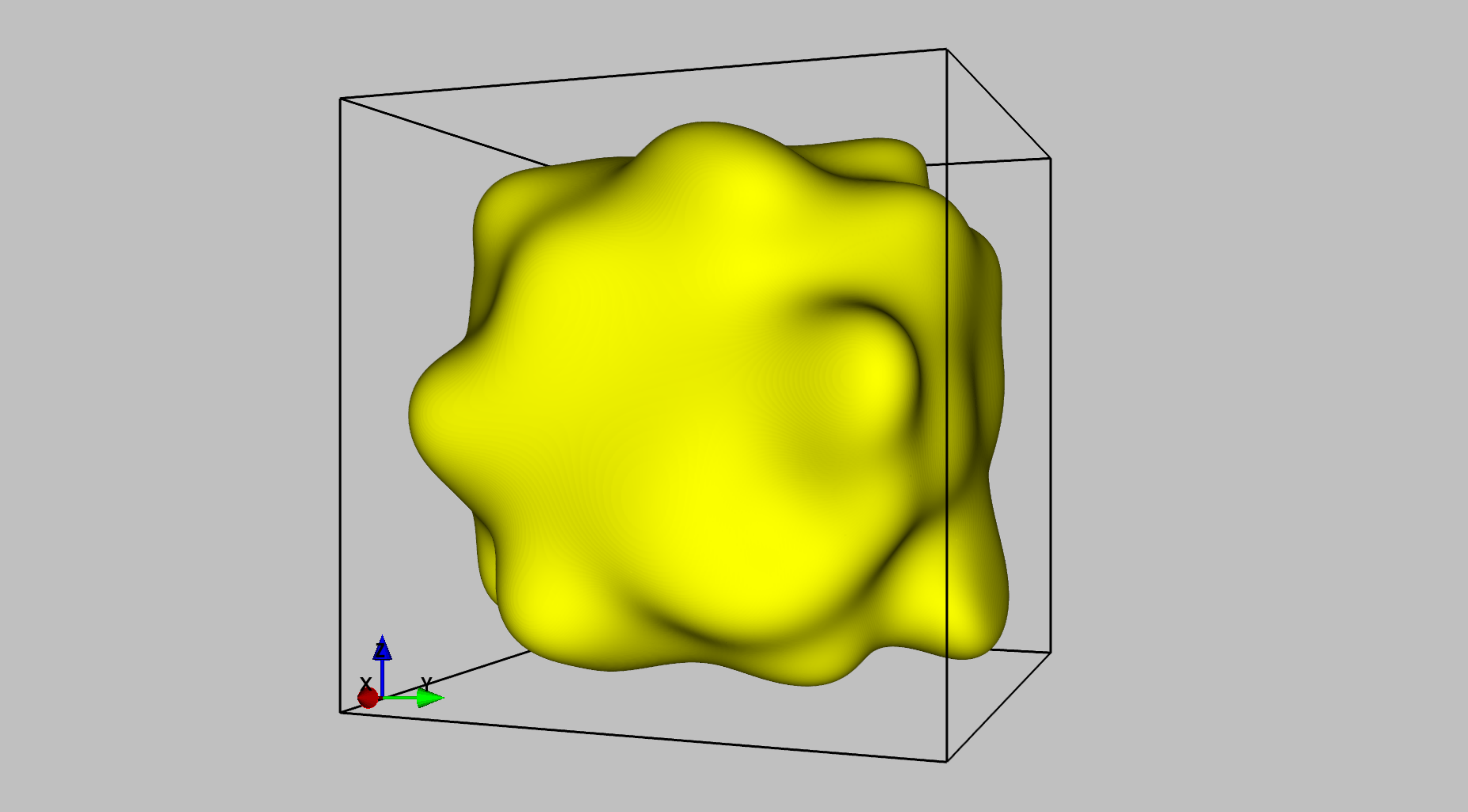}
\caption{The morphological instability of a precipitate growing from a very high supersaturation ($c_{\infty} = 0.7$) and hence at a very high velocity. The microstructures correspond to non-dimensional times of 1000 (top left), 1500 (top right), 2500 (bottom left) and 3500 (bottom right). The initial precipitate was spherical. The surface shown is iso-composition contour of 0.8. The systems size is 512 $\times$ 512 $\times$ 512.}\label{3DMorphInstability}
\end{figure}

In all the cases described above, the spherical precipitate remains spherical. However in Fig.~\ref{MorphInstability}, we show the microstructures (in a 2-D system) for a precipitate growing from a very high supersaturation ($c_{\infty} = 0.7$) for non-dimensional times of 2000, 4000, 6000, and 8000. In this case, the values of $L=6$ and $M=1$ are used (leading to $\beta = -0.367$ (that is a large and negative value). We have also used sustained noise (introduced once in every 10 steps) of strength about 0.5\%. As is clear from the figure, the growing precipitate does undergo morphological instabilities. Further, as is clearly seen from these figures, the morphologies are quite similar to those shown in~\cite{BrenerEtAl} for systems with very small interfacial anisotropy and very high undercooling; this is not surprising because in our case, the anisotropy is absent (isotropic interfacial energy) and the far-field composition is very high (large supersaturation) and hence has close similarities to the case discussed by Brener et al. For identical conditions ($c_{\infty} = 0.7$,$L=6$ and $M=1$), we have also seen morphological instabilities in 3-D -- see~\ref{3DMorphInstability}.

We believe that we are able to see morphological instabilities in these systems (and not in others for the following reason). The Gibbs-Thomson effect, since it increases the precipitate concentration above the equilibrium concentration, leads to a stabilising effect on the precipitate-matrix interface, However, when the kinetic coefficient if very large and negative (and when the normal velocity of the precipitate is high enough), it can reduce the $\Delta c$ and hence bring down the stabilising effect of the classical Gibbs-Thomson effect. Thus, the appropriate choice of $M$ and $L$, if it makes $\beta$ a large, negative value leads to morphological instabilities.

\section{Conclusions} \label{Section4}

\begin{itemize}

\item
The phase field modelling of growth kinetics of precipitates in 2- and 3-D systems shows that the precipitates grow faster in 3-D as compared to 2-D (as predicted by Zener and Frank);

\item
Purely due to the geometric effect (the Gibbs-Thomson effect in 3-D is twice that in 2-D in our case), the deviations from the classical Zener and Frank results are more in 3-D as compared to 2-D;

\item
At lower supersaturations, the precipitate composition is closer to that predicted by Gibbs-Thomson; this, in turn, leads to deviations from the growth rate predicted by Zener and Frank (who neglected the Gibbs-Thomson effect); 

\item
We have found that the growth rates in systems with non-constant diffusivity are much closer to those predicted by ZF theory as compared to the same for systems with constant diffusivity; this can be attributed to to the relatively lower interfacial energy and the effect of a negative kinetic coefficient;

\item
While using Model C, it is important to choose the appropriate values of $M$ and $L$; for example, by appropriate choice of the kinetic parameters ($M$ and $L$) the kinetic coefficient $\beta$ in the generalised Gibbs-Thomson equation can be made negligible or take large, negative values; and,

\item
The precipitate-matrix interface undergoes morphological instabilities if the far-field composition is very high and the kinetic coefficient $\beta$ is large in magnitude and negative in sign. 

\end{itemize}

\section*{Acknowledgements}

We thank R. Sanakrasubramanian, DMRL and S. Chatterjee, IIT-Hyderabad for useful discussions; we thank IRCC, IIT-Bombay for funding through the grant 09IRCC16; one of us (AR) would like to thank GM for funding through 11GMTC002.

\bibliography{ArijitGuru}

\end{document}